\documentstyle[multicol,aps,epsf]{revtex}
\begin{document}
\title{Slow relaxation in granular compaction}
\author{E.~Ben-Naim, J.~B.~Knight, and E.~R.~Nowak} 
\address{The James Franck Institute, The University of Chicago, 
Chicago, IL 60637}
\maketitle
\begin{abstract}
Experimental studies show that the density of a vibrated granular
material evolves from a low density initial state into a higher
density final steady state. The relaxation towards the final density
value follows an inverse logarithmic law. We propose a simple
stochastic adsorption-desorption process which captures the essential
mechanism underlying this remarkably slow relaxation.  As the
system approaches its final state, a growing number of beads have to
be rearranged to enable a local density increase. In one dimension,
this number grows as $N=\rho/(1-\rho)$, and the density increase rate
is drastically reduced by a factor $e^{-N}$.  Consequently, a
logarithmically slow approach to the final state is found
$\rho_{\infty}-\rho(t)\cong 1/\ln t$.
\end{abstract}
\vspace{.2in}
\begin{multicols}{2}

Systems consisting of many macroscopic particles such as sand and
powder exhibit complex behavior despite their apparent
simplicity\cite{Jaeger}.  Shaking sand may result in size segregation,
rich pattern formation\cite{Evesque,Melo}, or creation of convection
rolls\cite{Ehrichs}.  Adding a few grains to a sand pile leads to
avalanches\cite{Bak,Jaeger1,Frette}. This wealth of phenomena has
stimulated analogies to other intriguing physical problems including
the motion of flux lines in superconductors, metastability in glassy
systems, and surface instabilities in fluids\cite{Jaeger}.

Despite growing interest, the physical principles underlying sand
remain largely unknown. Although individual grains are solid, it is
inappropriate to classify their collective properties as entirely
solid-like or liquid-like. Conventional thermodynamic theory is not
applicable to sand as thermal fluctuations are negligible,
$k_B T=0$.  Most of the experimental and the theoretical studies have
focused on spatial inhomogeneities of one sort or another, and much
less attention has been given to equally fundamental but simpler
situations such as compaction. Granular compaction can be viewed
as a model system for non-thermal relaxation in a disordered medium.

Granular compaction is relevant to production, packing, and
transportation of a wide array of industrial and agricultural products
such as food, grains, chemicals, drugs, and building materials.  A
granular assembly provides us with a practically uniform system where
upon vibration, the well-defined bulk density evolves from a loosely
packed mechanically stable initial state into a denser final state
(see ref \cite{Knight} and references therein).  The system exhibits
no convection, spatial patterns, or oscillatory behavior.  The system
slowly explores available microscopic configurations, and eventually
low-density metastable configurations are eliminated.  Several
mechanisms have been proposed to explain the kinetics of compaction.
The concept of compactivity, which quantifies how far the density is
from its maximal value, was argued to play the same role as temperature
does in conventional statistical
mechanics\cite{Edwards,Mehta,Mehta1}. Such a theory predicts fast
exponential relaxation of the density with two relaxation times
associated with collective and individual
modes\cite{Barker}. According to another theory, the motion of the
voids filling the space between the particles is effectively
diffusive, and as a result a power-law relaxation is
predicted\cite{Caram,Hong}. Although the proposed mechanisms are
compelling, their quantitative predictions fail to describe the time
dependence observed experimentally \cite{Knight}.

In the compaction experiment of Knight et al\cite{Knight},
monodisperse 2mm diameter glass beads are confined to a 1.88cm
diameter, 1m long pyrex tube. The tube is tapped vertically by a
vibration exciter. The duration of each pulse is much smaller than the
waiting time between different taps so that all of the kinetic energy
of the beads is dissipated, and the beads are at rest before the next
tap. The column density was measured by a noninvasive capacitive
technique. The beads are initially loosely packed with a volume
fraction of $0.577\pm0.005$. Precautions were taken to prevent
humidity (the tube is evacuated), convection, and electrostatic
charging. The vibration intensity is parameterized by $\Gamma$, the
ratio of the peak acceleration of a tap to $g=9.81$ m/s$^2$. In
general, the steady-state density increased monotonically with
increasing $\Gamma$. In Fig 1, the time dependence of the packing
fraction, $\rho(t)$, is shown for a representative vibration intensity 
\cite{Knight}.  Shown also is a four-parameter fit 
$\rho(t)=\rho_f-\Delta\rho_{\infty}/[1+B\ln(1+t/\tau)]$. The
parameters $\rho_f$, $\Delta\rho_{\infty}$, $B$, and $\tau$ depend 
only on the acceleration $\Gamma$.  This inverse logarithmic form fits
the experimental data better than exponential or algebraic laws
suggested by the aforementioned compaction theories.  Logarithmic
relaxation has also been observed in the the decay of the slope of a
vibrated sandpile \cite{Jaeger-Liu}.

As the compaction progresses, individual grains move slowly, and when
a grain size void is created due to such motion, it is quickly filled
by a new grain. New grains can not move into space occupied by other
grains. In other words, they interact with their neighbors via a hard
core interaction. When the packing fraction is large, voids the size
of a particle are rare and a large number of voids have to be
rearranged to accommodate an additional particle and a local density
increase. Following this line of reasoning, we propose a useful
heuristic argument.  Let us denote the volume of a particle by $V$ and
the pore volume per particle by $V_0$. Thus the packing fraction is
$\rho=V/(V+V_0)$, or alternatively, $V_0=V(1-\rho)/\rho$. Assuming
that $N$ particles are rearranged in such a way that they contribute
their entire free volume to create a particle size void, $NV_0=V$.  We
find $N=\rho/(1-\rho)$. The time associated with such a rearrangement
should increase exponentially with $N$, $T\sim e^N$.  Consequently,
the density increases according to the following rate equation
$d\rho/dt\propto e^{-N}=e^{-\rho/(1-\rho)}$.  The solution of this
equation is given asymptotically by $\rho(t)\cong 1-1/\ln t$.  We made
the unrealistic assumption that all of the free volume is available
for large void creation. Furthermore, we ignored the structure of the
granular assembly, and the fact that the density cannot exceed the
close-packed value. These features can be avoided by considering the
one-dimensional situation where a void can be uniquely ``assigned'' to
its neighboring particle.  Interestingly, the exponential reduction
factor agrees with analytic results in 1D.

\begin{figure}\vspace{-.2in}
\centerline{\epsfxsize=9cm \epsfbox{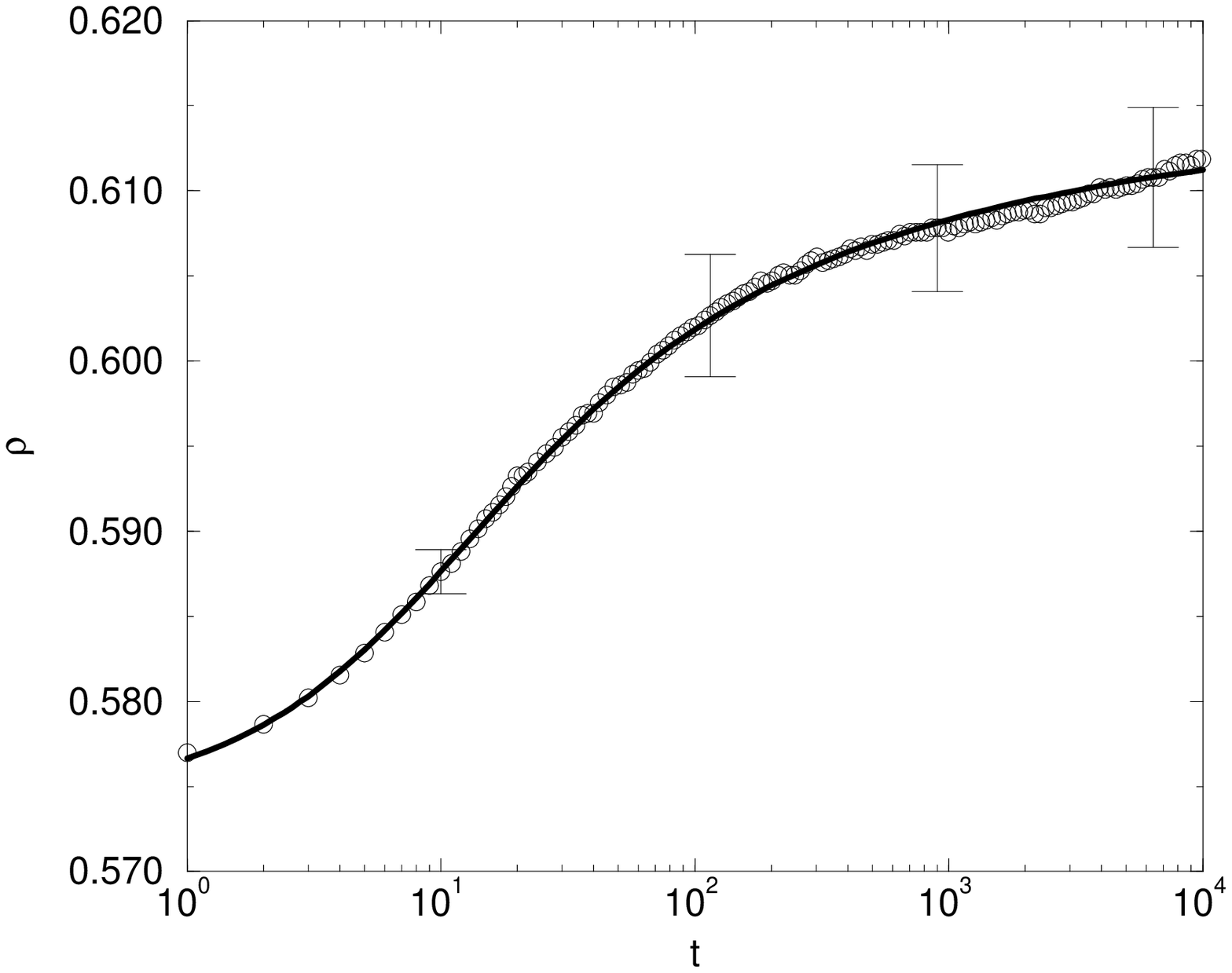}}
\vspace{-.2in}
\label{fig1}
{\small Packing fraction vs. time. Compaction data ($\circ$) near the
bottom of the tube for acceleration $\Gamma=2.3$. The data represent
an average over 5 different runs and the error bars correspond to the
rms variations between runs.  The solid line is a least square fit to
the inverse logarithmic form discussed in the text.}
\end{figure}

We now propose a simple adsorption-desorption theory for granular
compaction. The corresponding stochastic process was previously
studied in the context of chemisorption
\cite{Evans,Stinchcombe,Krapivsky} and protein binding\cite{McGhee}.
The model is defined as follows: Identical spherical particles of unit
diameter adsorb uniformly from the bulk to a substrate with rate $k_+$
and desorb with rate $k_-$.  In other words, $k_+$ adsorption attempts
are made per unit time per unit length, and similarly, the probability
that an adsorbed particle desorbs in an infinitesimal time interval
$(t,t+dt)$ is $k_-dt$.  While the desorption process is unrestricted,
the adsorption process is subject to excluded volume constraints, {\it
i.e.}, particles can not adsorb on top of previously adsorbed
particles.  The attempted adsorption event in Fig.~2 is thus rejected.
This stochastic process is well-defined in arbitrary
dimension. However, we restrict our attention to one-dimension where
analytic results are available.  This ``car-parking'' process clearly
satisfies detailed balance, and after a sufficiently long time, the
system reaches equilibrium.  The proposed model can be viewed as a
homogeneous one-dimensional gas of hard spheres \cite{Tonks}.

To analyze the parking model, it is useful to consider first the
simpler lattice version of the process where particles occupy a
single lattice site. In this case the density, $\rho$, satisfies the
Langmuir mean-field equation $d\rho/dt=k_+(1-\rho)-k_-\rho$. The gain
term is proportional to the fraction of unoccupied space, while the
loss term is proportional to the density itself. The steady-state 
density, $\rho_{\infty}$, which is obtained by imposing $d\rho/dt=0$,
depends on the rate ratio $k\equiv k_+/k_-$ only,
$\rho_{\infty}=k/(1+k)$. The system relaxes exponentially towards its
final state,
$\rho_{\infty}-\rho(t)=(\rho_{\infty}-\rho_0)e^{-t/\tau}$, with
$\rho_0$ the initial density. The relaxation rate is simply a sum of
the adsorption rate and the desorption rate $\tau^{-1}=k_++k_-$.

\begin{figure}\vspace{.2in}
\centerline{\epsfxsize=7.5cm \epsfbox{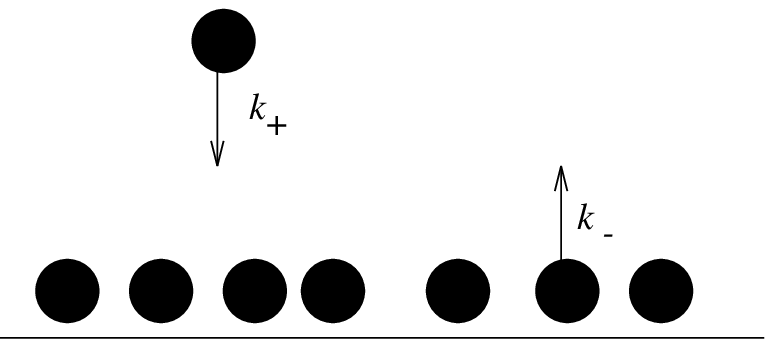}}
\vspace{.2in}
\label{fig2}
Fig 2 {\small The adsorption-desorption process.} 
\end{figure}

On the continuum, however, only a fraction of the empty space is
available for adsorption.  It was previously shown that in equilibrium
the probability $s(\rho)$ that an adsorption event is successful
equals $e^{-\rho/(1-\rho)}$\cite{Krapivsky}.  This so-called
``sticking coefficient'' equals unity in the low density limit, and
the above mean-field treatment holds. On the other hand, in the high
density limit this coefficient vanishes exponentially, $s(\rho)\propto
e^{-1/(1-\rho)}$, as $\rho\to 1$.  The excluded volume interaction
effectively reduces the adsorption rate, $k_+\to k_+(\rho)=
k_+s(\rho)$.  Thus, a modified Langmuir equation can be written
\begin{equation}
{d\rho\over dt}=k_+(1-\rho)e^{-{\rho/(1-\rho)}}-k_-\rho.
\label{rateq}
\end{equation}
This evolution equation was constructed to give the exact equilibrium
density, and is reasonable as long as the system is sufficiently
close to equilibrium.

Using Eq.~(\ref{rateq}), the equilibrium density is obtained from the
following transcendental equation, $\alpha e^{\alpha}=k$, with
$\alpha=\rho_{\infty}/(1-\rho_{\infty})$.  We find the following 
leading behavior in the two limiting cases
\begin{equation}
\rho_{\infty}(k)\cong\cases{k&$k\ll1$;\cr1-(\ln k)^{-1}&$k\gg1$.\cr}
\label{rhoeq}
\end{equation}
While the mean-field linear relation is recovered in the dilute limit,
the dense limit is characterized by a logarithmic cusp.  By contrast,
mean-field predicts a power-law dependence of the density
$\rho_{\infty}\cong1-1/k$, for $k\gg 1$. The effect of the volume
exclusion constraint is striking, a huge adsorption to desorption rate
ratio, $k\cong10^9$, is necessary to achieve a $0.95$ steady-state
occupancy.  It is also interesting to consider the equilibrium void
distribution. The density of voids of size $x$ is exponential
$P_{\infty}(x)=\beta e^{-\alpha x}$, with 
$\beta=\rho_{\infty}^2/(1-\rho_{\infty})$ and the previously defined
$\alpha$ \cite{Krapivsky}. Indeed, the sticking coefficient is
proportional to the density of voids larger than a particle size,
$s(\rho)\propto\int_1^{\infty}\, dx P_{\infty}(x)$.

We now focus on the relaxation properties of the system.  The granular
compaction process corresponds to the high density limit, and we thus
focus on the desorption-controlled case, $k\gg1$.  Hence, let us fix
$k_+$ and consider the limit $k_-\to 0$ of Eq.~(\ref{rateq}).  The
early time behavior is dominated by adsorption and can be obtained by
neglecting the desorption term.  For simplicity, we consider a
vanishing initial concentration. For sufficiently early times, $t\ll
1/k_+$, the density grows linearly in time, $\rho(t)\cong k_+t$. At
the later stages of the process, $t\gg 1/k_+$, the system approaches
complete coverage, $\rho_{\infty}=1$, according to
\cite{Krapivsky,Jin}
\begin{equation}
\rho(t)\cong \rho_{\infty}-{1\over\ln (k_+t)}. 
\label{rhot}
\end{equation}
This is confirmed by numerical simulations in one dimension.  Use of
Eq.~(\ref{rateq}) is justified {\it a posteriori} since the system
evolves slowly and has enough time to equilibrate.  The inverse
logarithmic behavior is simply a reflection of the exponentially
suppressed adsorption in the dense limit. In writing
Eq.~(\ref{rhot}) we neglected higher order logarithmic corrections
such as $\ln \ln (k_+t)$. We conclude that the excluded volume
interaction gives rise to a slow relaxation.

Eq.~(\ref{rhot}) holds indefinitely only for the truly irreversible
limit of the parking process, {\it i.e.}, for $k=\infty$. For large
but finite rate ratios, the final density is given by
Eq.~(\ref{rhoeq}).  As the density approaches this steady state value,
the loss term becomes significant and should be taken into account.
The crossover time between the two different relaxation regimes,
$t_0$, can be conveniently estimated by equating the time dependent
density of Eq.~(\ref{rhot}) with the equilibrium density of
Eq.(\ref{rhoeq}) $1-1/\ln(k_+t_0)=1-1/\ln(k_+/k_-)$, and as a result
$t_0\cong1/k_-$.  For $t\gg t_0$, the loss term is no longer
negligible.  By computing how a small perturbation from the steady
state decays with time, an exponential relaxation towards the steady
state is found $|\rho_{\infty}-\rho(t)|\propto e^{-t/\tau}$ for $t\gg
t_0$.  The relaxation time is indeed $t_0$, however, an additional
logarithmic correction occurs, $\tau=t_0(1-\rho_{\infty})^2\simeq
t_0/(\ln k)^2$. The above results can be simply understood: the early
time behavior of the system follows the irreversible limit of
$k_-=0$. Once the system is sufficiently close to the steady-state,
the density relaxes exponentially to its final value (Fig 
3). Hence, two relaxation curves with sufficiently large final
densities are indistinguishable over a significant temporal range.

\begin{figure}
\label{fig3}
\vspace{-.2in}
\centerline{\epsfxsize=9.5cm \epsfbox{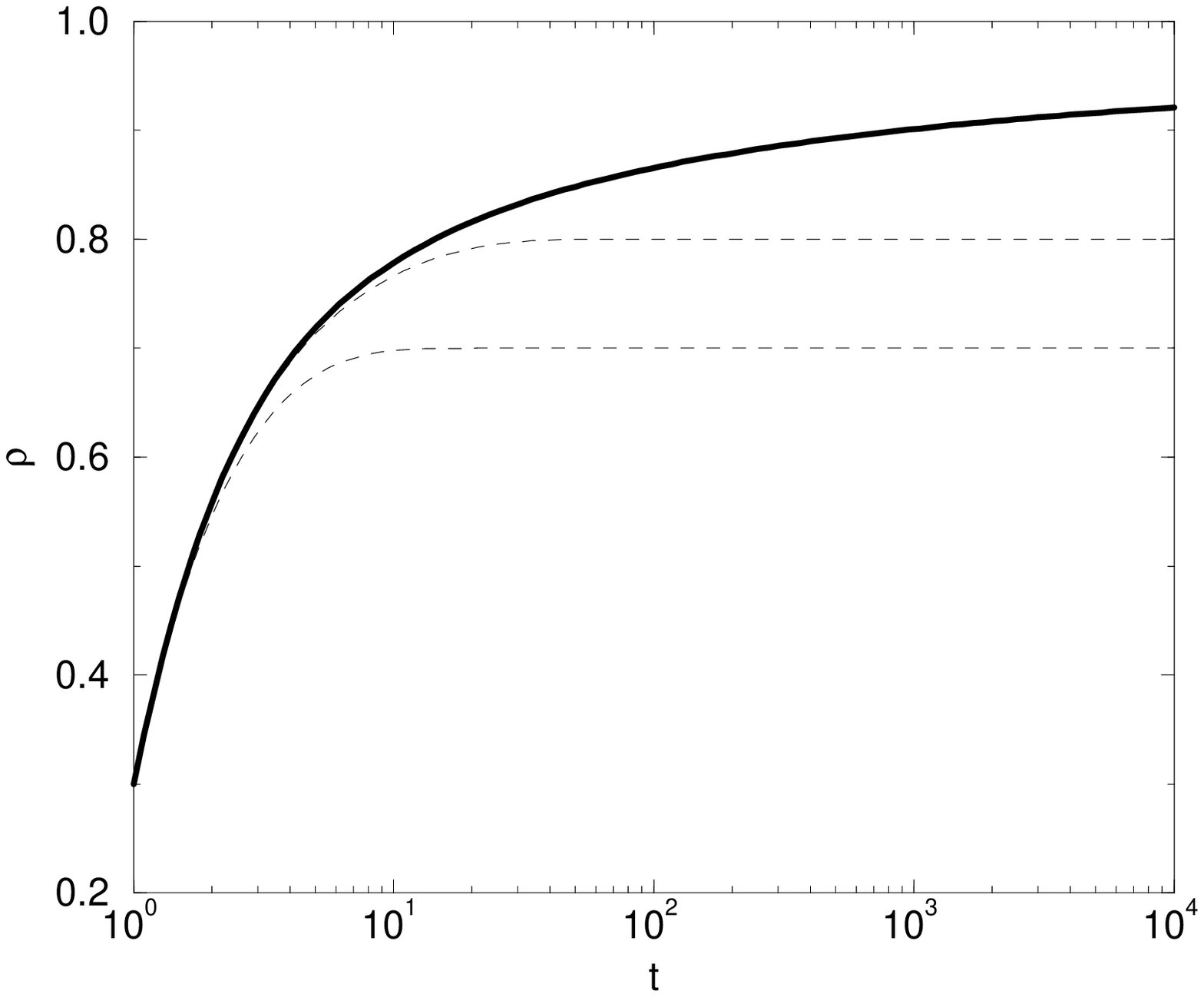}}
\vspace{-.2in}
Fig 3 {\small The coverage versus time. Shown are solutions of Eq.~(1) with
$k_+=1$. The upper curve corresponds to the irreversible limit
$\rho_{\infty}=1$, while the lower two curves represent reversible
parking with equilibrium densities of $\rho_{\infty}=0.7$ and $0.8$
respectively.}
\end{figure}

In the desorption-controlled limit, the void left by a desorbed
particle is quickly filled by an adsorbed particle.  Hence, the
desorption mechanism can be seen as playing a similar role to
diffusion of particles. Indeed, theoretical studies show that
adsorption processes where adsorbed particles diffuse rather than
desorb give rise to the same density relaxation as in
Eq.~(\ref{rhot})\cite{Privman}.  Our treatment was restricted to
one-dimensional processes, but we expect that the results hold in
higher-dimension as well. Hence, the above relaxation is rather robust
and insensitive to many microscopic details.

The inverse logarithmic density relaxation towards the steady-state is
the same one observed for granular compaction (see Fig 1).
Furthermore, the experimentally observed relaxation curves which
correspond to large compaction, $\Gamma>2.7$, are indistinguishable
over the observed time range consistent with the above theory. The
largest packing fraction observed in the experiment are well below the
maximal close-pack value in 3D. Thus, according to the parking theory
the logarithmic relaxation regime may be followed by an exponential
relaxation. There exists a maximal time scale which characterizes the
temporal behavior. Such a prediction is consistent with the
experimental observation that steady state is achieved within a finite
time. Finally, once the system reaches its final state,
density fluctuations can be measured and compared with Monte Carlo
simulations of the parking process. Our preliminary results indicate
that measured and simulated Power Spectrum Density (PSD) of the
density fluctuations are strikingly similar \cite{Nowak}. The high
frequency limit is Lorentzian ($f^{-2}$), while the low density limit
is white-noise ($f^0$). The intermediate regime is approximately a
powerlaw $f^{-\alpha}$, with $0.7<\alpha<1.4$.

In a realistic granular material, an individual particle can not
penetrate its neighbors, but it also must be in contact with several
other particles.  Our model properly accounts for the hard core
repulsion, but it ignores mechanical stability. We argue that
in the long time limit mechanical stability can not play a significant
role in determining the motion of individual grains during the
compaction process. Instead, the motion is limited primarily by the
presence of other beads.  The tradeoff, however, is that such a simple
theory can not make predictions about the final density.
Nevertheless, it does elucidate the leading mechanism in granular
compaction.  It will be very useful to try and incorporate the
important distinction between total void space and the available void
space into the recent theories concerning compaction and packing
\cite{Edwards,Hong,Herrmann}.

In conclusion, we have studied theoretically density relaxation
towards steady state in granular compaction using a {\it microscopic}
model in one dimension. Due to volume exclusion, exponentially growing
time scales are associated with cooperative motion of grains. As a
result the approach towards the steady state is an inverse logarithmic
one.  Since the argument leading to the logarithmic relaxation is a
general one, the results should hold in a large class of physical
situations, for example when the shaking is horizontal rather the
vertical or when the grains are aspherical.  Interestingly, the same
mechanism is responsible for the long time necessary for packing
grains in a bowl, parking a car in a busy street, or even entering a
crowded room.

We are grateful to H.~M.~Jaeger and S.~R.~Nagel for many stimulating
discussions.  We also thank P.~Dimon, H.~J.~Herrmann, D.~C.~Hong,
L.~P.~Kadanoff, P.~L.~Krapivsky, J.~F.~Marko, A. Mehta, T.~A.~Witten
for useful discussions.  This work was supported by The MRSEC Program
of the National Science Foundation under Award Number DMR-9400379, and
by the Department of Energy Grant No. DE-FG02-92ER25119.  We also wish
to acknowledge support from NSF grant \#92-08527 (EB) and NSF (JBK).

\end{multicols}
\end{document}